\documentclass[superscriptaddress,reprint,amsmath,amssymb,aps,prb]{revtex4-2}

\usepackage{graphicx}
\usepackage{dcolumn}
\usepackage{bm}
\usepackage{physics}
\usepackage{qcircuit}
\usepackage{hyperref}

\usepackage{color,soul}

\begin{document}

\title{Variational quantum eigensolver for frustrated quantum systems}

\author{Alexey Uvarov}\homepage{http://quantum.skoltech.ru}
\email{alexey.uvarov@skoltech.ru}
\affiliation{Skolkovo Institute of Science and Technology, Moscow 121205, Russia}
\author{Jacob D.~Biamonte}
\affiliation{Skolkovo Institute of Science and Technology, Moscow 121205, Russia}
\author{Dmitry Yudin}
\affiliation{Skolkovo Institute of Science and Technology, Moscow 121205, Russia}

\date{\today}

\begin{abstract}
Hybrid quantum-classical algorithms have been proposed as a potentially viable application of quantum computers. A particular example---the variational quantum eigensolver, or VQE---is designed to determine a global minimum in an energy landscape specified by a quantum Hamiltonian, which makes it appealing for the needs of quantum chemistry. Experimental realizations have been reported in recent years and theoretical estimates of its efficiency are a subject of intense effort. Here we consider the performance of the VQE technique for a Hubbard-like model describing a one-dimensional chain of fermions with competing nearest- and next-nearest-neighbor interactions. We find that recovering the VQE solution allows one to obtain the correlation function of the ground state consistent with the exact result. We also study the barren plateau phenomenon for the Hamiltonian in question and find that the severity of this effect depends on the encoding of fermions to qubits. Our results are consistent with the current knowledge about the barren plateaus in quantum optimization. 

\end{abstract}

\maketitle

\section{Introduction} 
Based on effective dimensionality reduction of unsorted datasets and the ability to efficiently recognize patterns, neural network algorithms allow one to directly address properties of physical systems with no prior knowledge of the structure of their states. Indeed, with recent advances in machine learning simulation, relatively large spin Hamiltonians of up to a few hundreds entities might be approached \cite{Carleo2019}. Nevertheless, studying the properties of highly frustrated systems remains a highly nontrivial computational task \cite{Westerhout2020}. At the same time, quantum algorithms have several similarities with various traditional machine learning models. While recent times have seen rapid development in the use of quantum circuits as variational models of quantum machine learning \cite{Biamonte2017}, these techniques are still being developed to apply to the simulation of physical systems. 

Using variational quantum algorithms is widely considered as a promising approach towards practical application of quantum computers \cite{mcclean_theory_2016,Li2017,Li2017a,Temme2017,colless_computation_2018,Babbush2018,Kivlichan2018,barkoutsos_quantum_2018,Macridin2018,Klco2018,Setia2018,endo_variational_2019,grimsley_adaptive_2019,Ganzhorn2019,Parrish2019,Demers2019,Takeshita2020,Lubasch2020,akshay_reachability_2020}. The idea of such algorithms is to delegate as many calculations as possible to a classical device, thereby minimizing quantum resources. A particular example, the variational quantum eigensolver (VQE), represents an implementation of variational quantum circuits which uses a quantum computer to prepare a family of states characterized by a polynomial number of parameters and minimizes the expectation value of a given Hamiltonian within this family. This approach is among the first realized on small-  and mid-sized quantum computers \cite{peruzzo_variational_2014,omalley_scalable_2016,colless_computation_2018,li_variational_2019}. Thus, in  VQE one aims at finding the ground state of a quantum Hamiltonian with the use of a certain tunable ansatz state that is easy to prepare on a quantum device but hard to store in classical memory. Using the VQE one can evaluate at least the upper bound of lowest eigenvalue of a given Hamiltonian. 

Potential applications of the VQE include quantum chemistry \cite{cao_quantum_2018,kandala_hardware-efficient_2017,colless_computation_2018,nam_ground-state_2019,hempel_quantum_2018,grimsley_adaptive_2019}, condensed matter physics \cite{cade_strategies_2019,grant_initialization_2019,endo_calculation_2019,bravo-prieto_scaling_2020}, and lattice quantum field theory \cite{kokail_self-verifying_2019}, although discrete optimization is also within the scope of variational quantum algorithms \cite{farhi_quantum_2014,nannicini_performance_2019}. The hopeful advantage over classical computation lies largely in the ability of a quantum processor to manipulate the quantum states to avoid an exponentially large classical description. Thus, VQE presents a way to deal with quantum systems outside the range of classical computers. In general, optimization of a local quantum Hamiltonian is a complete problem for complexity class QMA, meaning that particular problems are out of reach even for a quantum computer. However, it is hoped that practical problems can be solved to acceptable tolerance. 

In condensed matter physics, frustrated systems with inhomogeneous interactions are hard to analyze owing to extra degrees of freedom to show up. On one hand, strong electron-electron interactions precludes perturbative expansion over single-electron wave functions. On the other hand, more advanced numerical approaches to strongly correlated systems, e.g., based on dynamical mean-field theory, treat the systems on a purely local manner. Whether modern quantum algorithms can give an edge in analyzing these models is an intensely studied question \cite{cade_strategies_2019,rungger_dynamical_2019,jaderberg_minimum_2020}. In this work, we analyze the performance of VQE for a one-dimensional model of spinless electrons with nearest- and next-nearest-neighbor interactions. This model represents a simple theoretical testbed to explore the physical properties of frustrated systems. Resulting from the competition between two types of interactions, a metallic state emerges even for strongly interacting systems. Interestingly, results of numerical simulations for finite size clusters unambiguously reveal that the ground state does not belong to the Luttinger liquid universality class \cite{Zhuravlev1997,zhuravlev_breakdown_2000,zhuravlev_one-dimensional_2001,Hohenadler2012,Karrasch2012}. 

Moreover, we also address a bottleneck of hybrid algorithms in the regard of the considered model, i.e., we inspect the onset of the so-called plateau regime. One of the more recently identified hurdles when implementing VQE is the so-called {\it barren plateau} effect \cite{mcclean_barren_2018}. The idea is that the variation of the gradient of any useful cost function decays exponentially with the number of qubits, provided that the quantum circuit used as ansatz in the VQE is long enough to implement an approximate solution. For shorter quantum circuits, the onset of the plateau regime seems to depend on the geometric locality of the cost function \cite{cerezo_cost-function-dependent_2020}. In VQE, the choice of cost function is motivated by physical problems in question, hence it typically has some kind of locality in its terms. Interestingly enough, the results for the latter are rather sensitive to fermion-to-qubit mapping as we discuss below.

\section{Numerical methodology} 
Assume that a quantum state $\ket{\psi} = \ket{\psi(\boldsymbol{\theta})}$ can be prepared on a quantum computer, with adjustable real parameters $\boldsymbol{\theta} = (\theta_1, ..., \theta_p)$. Then, by measuring each qubit in a local basis, one can measure expectation values of the type $\bra{\psi}\sigma_1\otimes ... \otimes \sigma_n\ket{\psi}$, where $\sigma_i$ are drawn from the set of Pauli matrices $\{I, X, Y, Z\}$. We will refer to such observables as Pauli strings. Any Hamiltonian acting on $n$ qubits can be decomposed into a sum of such Pauli strings, and hence we can evaluate the energy $\mathcal{E}(\boldsymbol{\theta})=\bra{\psi} H \ket{\psi}$ with respect to the state prepared on a quantum computer, as long as the number of Pauli strings is polynomial in $n$. The idea of the VQE is then to find the lowest energy state $\ket{\psi}$ among some set of states that can be prepared on the device. Most often, $\ket{\psi}$ is a state prepared by some quantum circuit with parametrized gates. In this case, the parameters are optimized in a classical computer using the quantum computer as a black box. Noteworthy that under some assumptions on the shape of the ansatz, one may also be able to access derivatives of the cost function \cite{mitarai_theory_2019,schuld_evaluating_2019,sweke_stochastic_2019,harrow_low-depth_2019}. 

For the numerical implementation in this study, we use the limited-memory  Broyden–Fletcher–Goldfarb–Shanno (L-BFGS) algorithm, which is a so-called quasi-Newton method. This method consists in approximately evaluating the Hessian matrix of the cost function followed by the step of the Newton method. We perform noise-free numerical simulations, on condition that the optimization algorithm has direct access to an exact value of energy, as though it was obtained with an infinite number of VQE shots. The gradients are estimated by finite differences. Each component of the gradient is estimated as $\partial_\theta\mathcal{E}(\theta_k) \approx (\mathcal{E}(\theta_k + \delta) - \mathcal{E}(\theta_k))/\delta$, where the parameter is chosen as $\delta = 10^{-6}$.

\subsection{Probe states} 
Clearly, the performance of VQE crucially depends on the choice of ansatz state. Most common approach uses the unitary version of the coupled cluster method, the unitary coupled cluster ansatz \cite{Shen2017,Dumitrescu2018}. For interacting spin problems, the (non)unitary coupled cluster ansatz can be composed out of spin flip operators \cite{Gotze2011}.  There is no direct evidence that this approach might be simulated on a classical computer in large scale, even when the series is truncated on low order terms \cite{Taube2006}. In principle, a quantum computer could efficiently prepare this state truncated up to some $k$th order using the Suzuki--Trotter decomposition \cite{Suzuki1976}. However, for a system of $n$ qubits it requires $\mathcal{O}(n^k)$ unitary gates, making this technique out of reach for contemporary quantum computers. Here we use a heuristic ansatz which consists of layers of parametrized two-qubit gates acting alternatively either on pairs $(1, 2), (3, 4), ..., (2k-1, 2k)$ or on pairs $(2, 3), (4, 5), ..., (2k, 2k+1)$. We will refer to this construction as a checkerboard ansatz \cite{Uvarov_Kardashin_Biamonte_2019}, although in other publications it is also known as alternating layered ansatz \cite{cerezo_cost-function-dependent_2020} or  parallel local circuit \cite{Brandao_Harrow_Horodecki_2016}. For an odd number of qubits, this construction is forced to skip one qubit in each layer. This is done in a cyclic manner, so that we skip the last qubit in the first layer, the first qubit in the second layer, the second qubit in the third layer, and so on.

The ansatz we use throughout the numerical simulations consists of typical two-qubit gates---see, e.g.,~\cite{barkoutsos_quantum_2018},

\begin{equation}
\label{eq:two-qubit_gate}
    U(\theta_1, \theta_2) = 
    \begin{pmatrix}
1 & 0 & 0 & 0 \\
0 & \cos{\theta_1} & e^{\imath \theta_2} \sin \theta_1 & 0 \\
0  & e^{-\imath \theta_2} \sin \theta_1  & -\cos{\theta_1} & 0  \\
0 & 0 & 0 & 1
\end{pmatrix}.
\end{equation}
This entangling gate has a convenient experimental implementation \cite{egger_entanglement_2019}. Moreover, the gate is universal, and hence, with access to single-qubit operations, any quantum circuit can in principle be realized on condition that the sequence is long enough. The gate further respects particle conservation, provided the so-called match gate layout is adhered to. When a fermion problem is cast into a spin problem using the Jordan--Wigner mapping, this construction conserves the total number of particles by keeping the $\ket{00}$ and $\ket{11}$ subspaces invariant. Of course, to start with the desired number of particles, one has to initialize the qubits by applying the Pauli $X$ gate. In principle, by adjusting the chemical potential, one already sets the number of particles seen in the ground state, but the ansatz restriction also excludes the exploration of the undesired states. In total, an ansatz of that form with $n$ qubits and $L$ layers has $2 L \lfloor n/2 \rfloor$ free parameters.

\subsection{Spin-fermion mapping} 
In fermionic Hamiltonians, one has to keep track of the canonical anticommutation relations. In contrast, qubits in a quantum computer are distinguishable particles that have no such relations. Therefore, in order to solve eigenvalue problems for a fermionic system on a quantum computer, one has to map one to another. One method to do that is known as the Jordan--Wigner transformation. Population-wise, the $j$th electron site is mapped to the $j$th qubit. The mapping of the fermionic operators to qubit operators, though, carries the parity information in the form of a phase multiplier. Thus, fermionic creation and annihilation operators $f^\dagger_j, \ f_j$ transform by the following rule:
\begin{equation}
\label{eq:jw}
    f^\dagger_j = \mathcal{Z}_j \otimes \sigma^+_j, \quad f_j = \mathcal{Z}_j \otimes \sigma^-_j,
\end{equation}
provided that $\mathcal{Z}_j=Z_1 \otimes .. \otimes Z_{j-1}$ and $\sigma^{\pm} = (X \pm iY)/2$. The order of site enumeration is important. For one-dimensional systems the procedure is clear enough, whereas for higher-dimensional systems one can enumerate the sites in a {\it snake} order, see, e.g., Ref.~\cite{cade_strategies_2019} and references therein. The advantage of this technique is found in the ease with which one can enforce the total particle number. Nevertheless, the locality of the operators changes dramatically in order to contain the parity information. Typically, local fermionic operators map to $n$-local spin operators.

Another version of spin-fermion mapping is provided by the Bravyi--Kitaev transformation \cite{bravyi_fermionic_2002}. In this method, the population and parity information are mixed in a procedure that allows the fermionic operators to map to $\log n$---local spin operators. However, under this transformation, there is no obvious way to enforce particle number conservation, except for adding extra penalty terms to the Hamiltonian.

\section{Results and discussion}
We apply the methodology described in the previous section to study the ground state properties of interacting quantum many-body systems. Consider the Hamiltonian of one-dimensional spinless fermions
\begin{equation}
\label{eq:hubbard_nnn}
    H = -t \sum_{\langle i,j\rangle}c^\dagger_i c_j + V_1\sum_in_i n_{i+1} + V_2\sum_in_i n_{i+2},
\end{equation}
where $c_i^\dagger$, $c_i$ stand for creation, annihilation operators of an electron at $i$th site with a number of $n_i=c_i^\dagger c_i$ electrons, $t$ is the hopping energy, $V_1$ and $V_2$ are nearest- and next-nearest-neighbor Coulomb interactions. In the first term, the summation $\langle\ldots\rangle$ is implied over nearest neighboring sites. In the following, we impose periodic boundary conditions. Earlier numerical studies suggest that when the $V_2$ is increased, on condition the ratio $V_1 / V_2=2$ is fixed, this model is characterized by appearance of a metallic state but does not belong to the Luttinger liquid class \cite{zhuravlev_breakdown_2000}. Away from this ratio, the model is gapped. For $V_2 = 0$, a regular interacting model of spinless fermions is recovered.

\begin{figure}
    \centering
    \includegraphics[width=\linewidth]{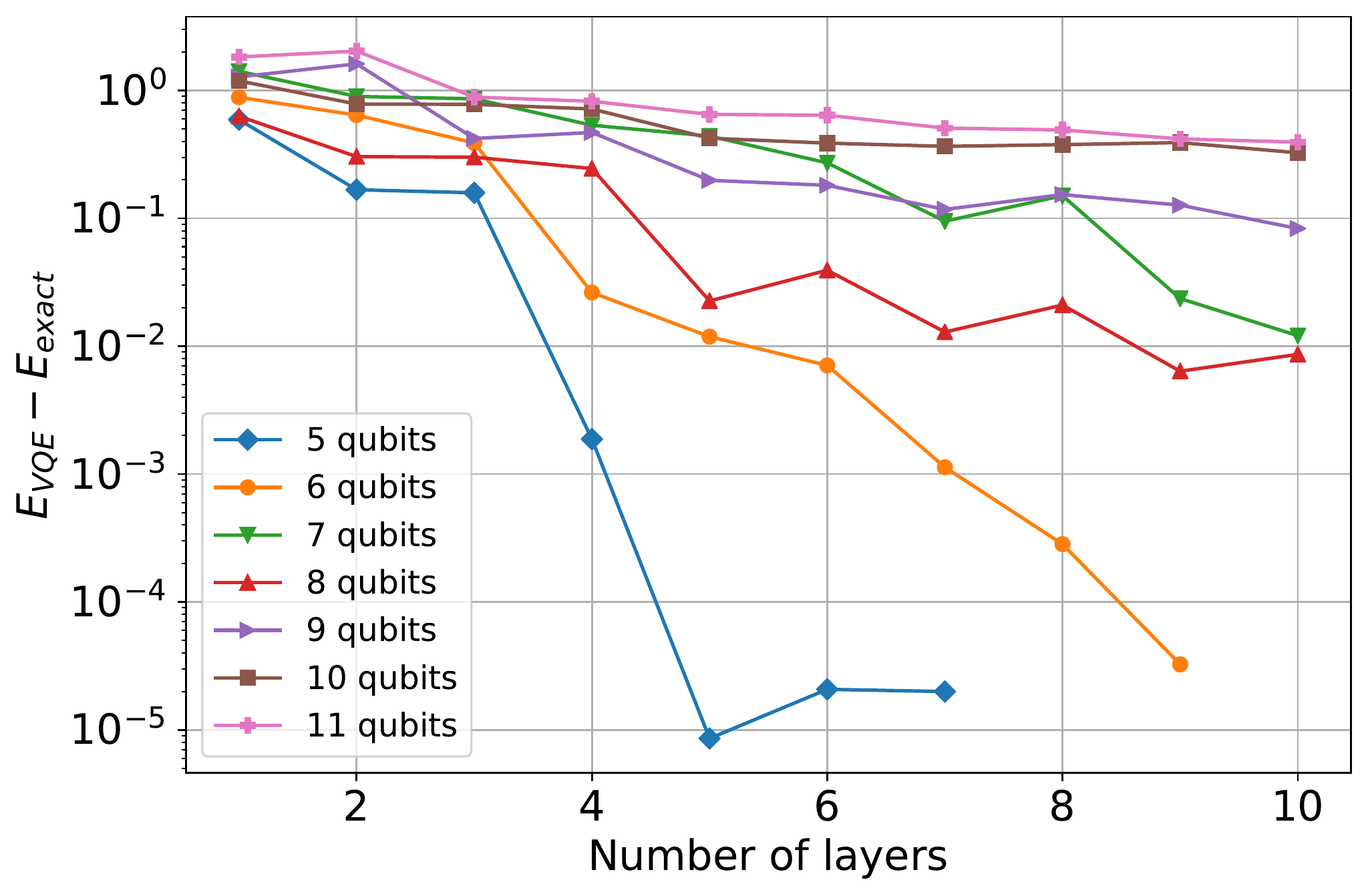}
    \caption{Convergence of the VQE energy to the true ground state energy for the model of Eq.~(\ref{eq:hubbard_nnn}) versus the number of layers, on condition $V_1 = 2t, V_2 = t$. Cases of $n \leq 4$ qubits are not shown as they converged to exact solution within two layers.}
    \label{fig:hubbard_nnn_solutions}
\end{figure}

We numerically test the difference between exact solution and the VQE solution for $V_2 = t, V_1 = 2t$ for various numbers of qubits and ansatz layers. For one layer, the parameters are instantiated as i.i.d.~random normal variables with zero mean and variance $0.1$. Then, when the second layer is appended, the existing parameters are kept at their previous optimal values, and the new parameters are instantiated as random variables with the same distribution. In particular, Fig.~\ref{fig:hubbard_nnn_solutions} illustrates how efficiently the total energy in the model with next-nearest-neighbor interactions is minimized by VQE, i.e., it shows the absolute error of the VQE solution relative to the exact one. Clearly, more layers are needed to achieve the same accuracy for more qubits, although the exact scaling remains unclear. Notably, for $n=11$ qubits, the energy error remains roughly the same despite the increase in the number of layers. For some tests, the error decay is not strictly monotonic with the number of layers. This may be caused by the choice of gate family in the ansatz. However, the error has the tendency to decrease in all cases, as expected. 

{}

\begin{figure}
    \centering
    \includegraphics[width=\linewidth]{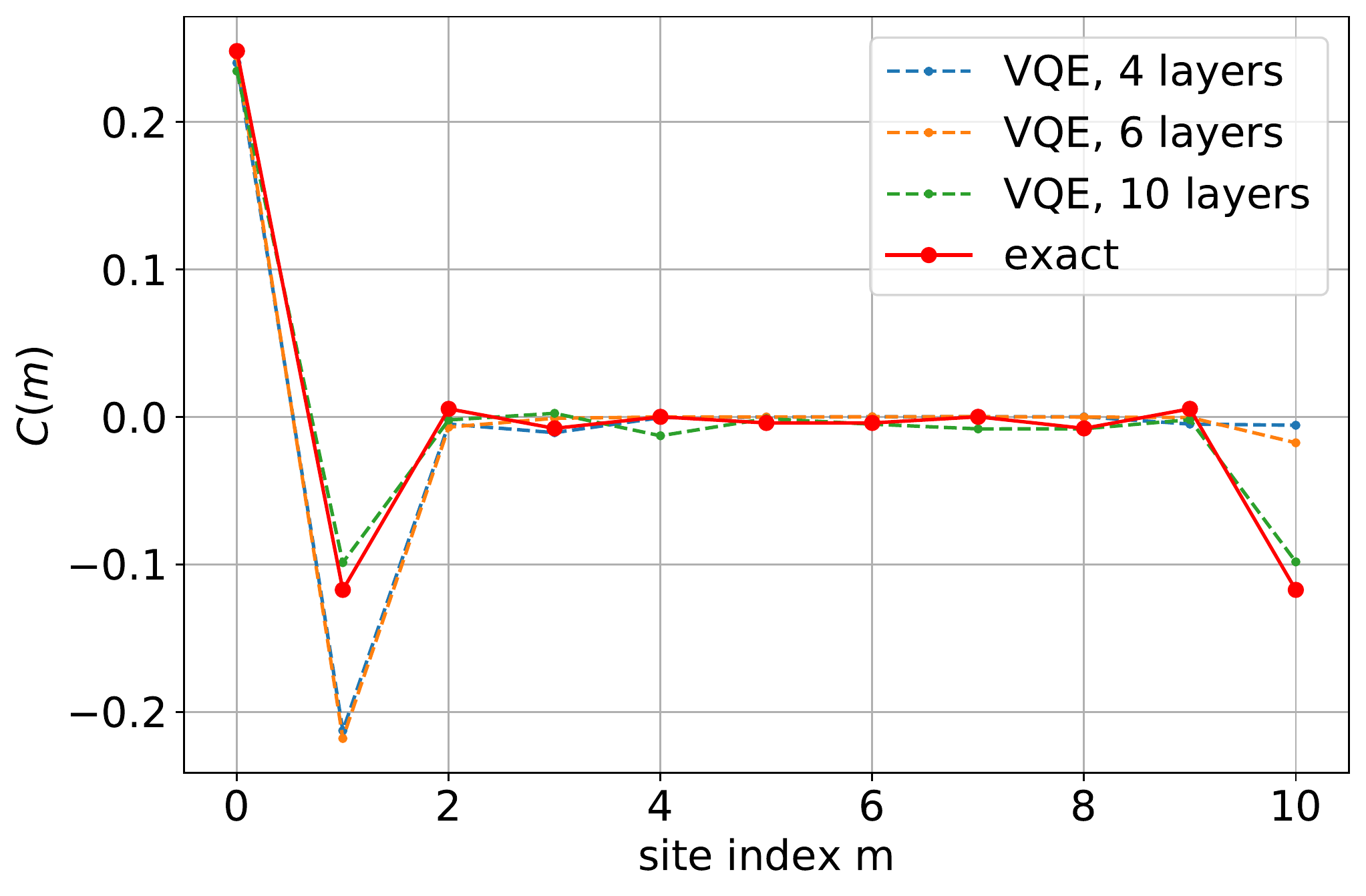}
    \caption{Density-density correlation function between spatially separated lattice sites. Solid line denotes the exact solution as obtained by virtue of exact diagonalization of the Hamiltonian (\ref{eq:hubbard_nnn}); dashed lines stand for the VQE solutions with different ansatz depths. Here $V_1 = 2t, V_2 = t$.}
    \label{fig:correlation}
\end{figure}{}

\subsection{Correlation function}
\begin{figure}
\includegraphics[width=\linewidth]{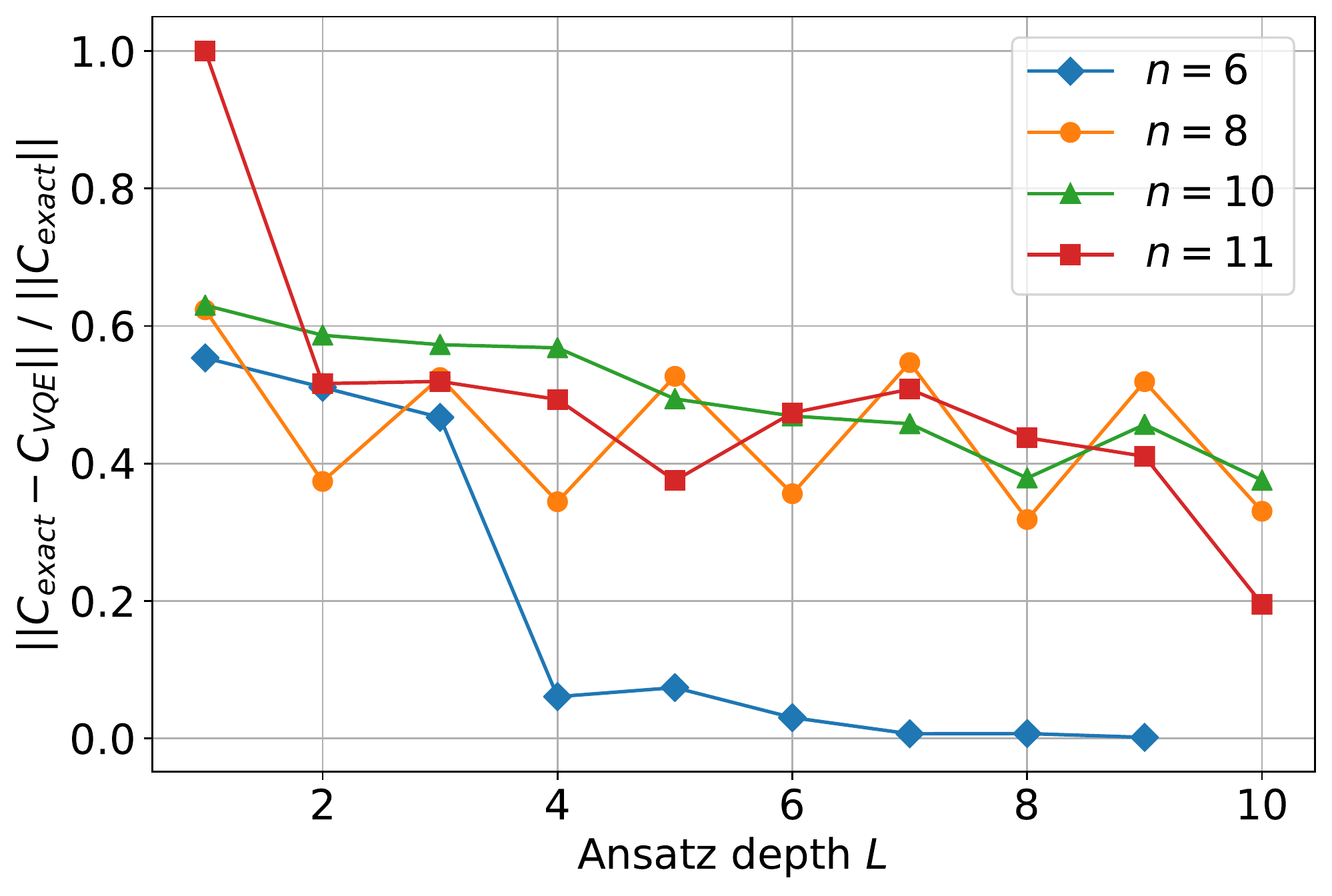}
\caption{Relative error of the correlation function as obtained with the VQE method, $C_{VQE}$, with respect to the exact solution, $C_\mathrm{exact}$, as a function of the number of layers, $L$.}
\label{fig:corr_errors}
\end{figure}

The energy error can be shown to be closely related to the infidelity between the true ground state and the variational approximation \cite{biamonte_universal_2019}. In other words, the error in energy is close to zero if the variational solution lies close to the ground state subspace. However, both of these metrics are useful only when we possess enough information on the exact solution. Since in real applications we want to find the true energy in the first place, we cannot assume this knowledge {\it a priori}. It is therefore also interesting to consider the convergence of other physically relevant observables. Specifically, we consider the convergence of the density-density correlation function $C(m) = \langle n_0 n_m\rangle - 
\langle n_0\rangle \langle n_m\rangle$. Note that the expectation values in this definition are taken explicitly with respect to the VQE wave function. Results of numerical simulations for $n=11$ qubits are shown in Fig.~\ref{fig:correlation}. We depict how the correlation function depends on the qubit number $i$ and the number of layers. The curves show qualitatively similar behavior even for small number of layers. However, a substantial decrease in error is observed only at eight layers in the ansatz. While finite size effects for $n=11$ qubits prevent us from making a reliable prediction about the functional behavior of the correlation function, we can nonetheless estimate the discrepancy between the correlation function of the exact solution and that obtained by the VQE. To provide a quantitative estimate we depict relative error of the correlation function as implemented in the VQE routine with respect to the exact solution in Fig.~\ref{fig:corr_errors}. It is worth mentioning that for a more complicated system the exact solution is, of course, unavailable. Nonetheless, the changes in the correlation function with increasing depth can potentially hint on the overall convergence.

{}

\subsection{Barren plateaus in VQE optimization} 
It was noticed in Ref.~\cite{mcclean_barren_2018} that, given a long enough parametrized ansatz, the gradients of any reasonable cost function will be exponentially small with respect to the number of qubits. Further on, it was discussed that for intermediate size quantum circuits, the onset of this {\it barren plateau} effect depends on the nature of the cost function \cite{cerezo_cost-function-dependent_2020}. In particular, two types of gradient behavior have been highlighted:

\begin{figure}
\includegraphics[width=\linewidth]{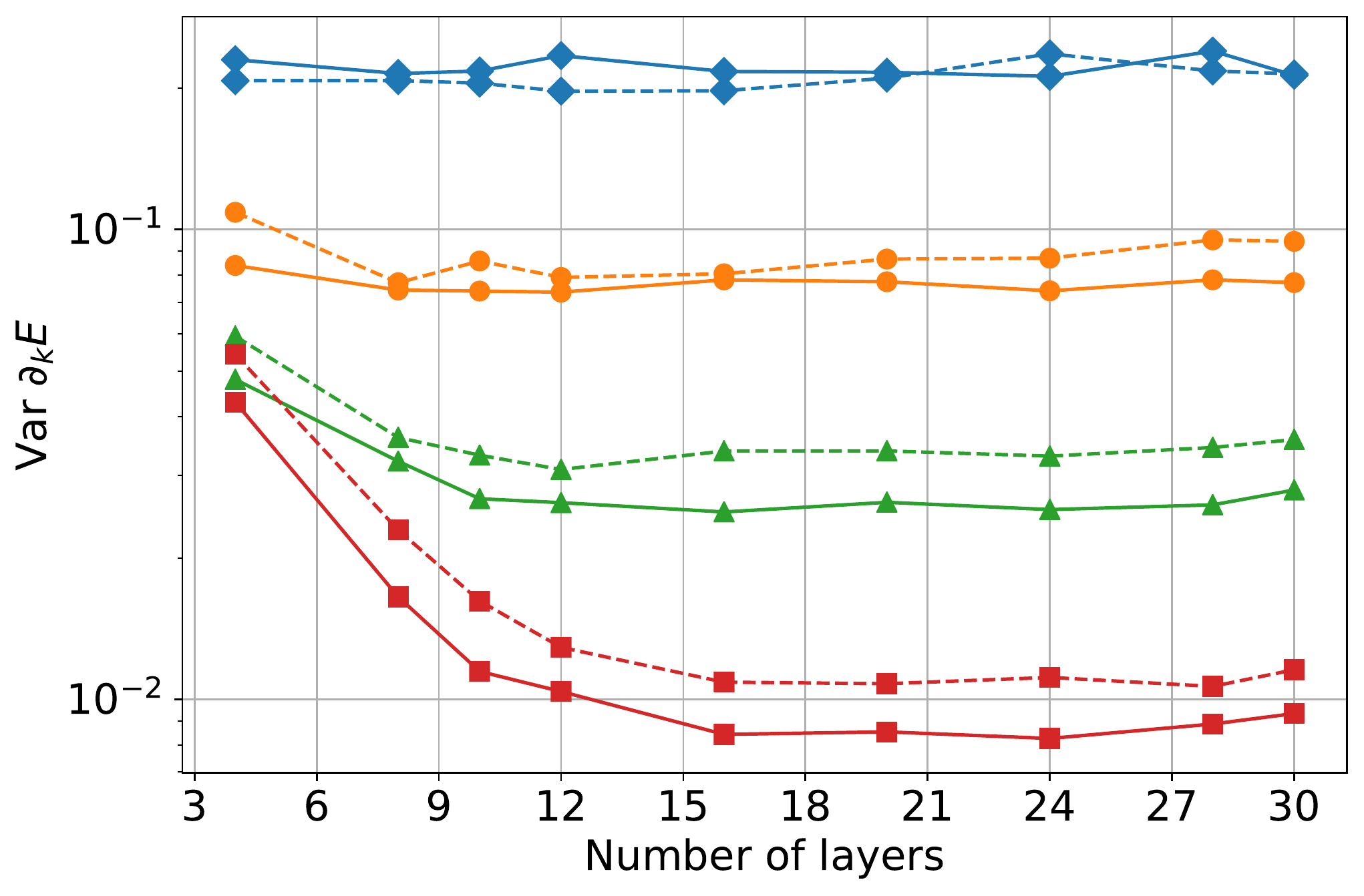}
\caption{Barren plateau effect for the Hamiltonian of Eq.~(\ref{eq:hubbard_nnn}) with $V_1=t$ and $V_2=0$ (dashed lines), as well as $V_1=2t$ and $V_2=t$ (solid lines) versus the number of qubits as realized by virtue of Jordan--Wigner mapping. Diamonds: four qubits; circles: six qubits; triangles: eight qubits; squares: 10 qubits. Markers in Figs.~\ref{fig:bk_hubbard} and \ref{fig:tfim} follow the same convention.}
\label{fig:jw_hubbard}
\end{figure}{}

\begin{enumerate}
    \item $\mathrm{Var} \ \partial_\theta\mathcal{E}(\boldsymbol{\theta}) \in \mathcal{O}(e^{-L})$, where $L$ is the number of ansatz layers. In this case, one can use a logarithmic number of layers without hitting the plateau.
    \item $\mathrm{Var} \ \partial_\theta\mathcal{E}(\boldsymbol{\theta}) \in \mathcal{O}(2^{-n})$ independent of the number of layers. In this case, the plateau is reached, thwarting VQE performance unless the initial conditions are somehow favorable.
\end{enumerate}{}
In the following, we examine how this effect takes places in the VQE process for the model of Eq.~(\ref{eq:hubbard_nnn}). To do that, we estimate the values of the gradient components in $N=50$ random points in the parameter space for some values of depths and qubit numbers.

\begin{figure}
\centering
\includegraphics[width=\linewidth]{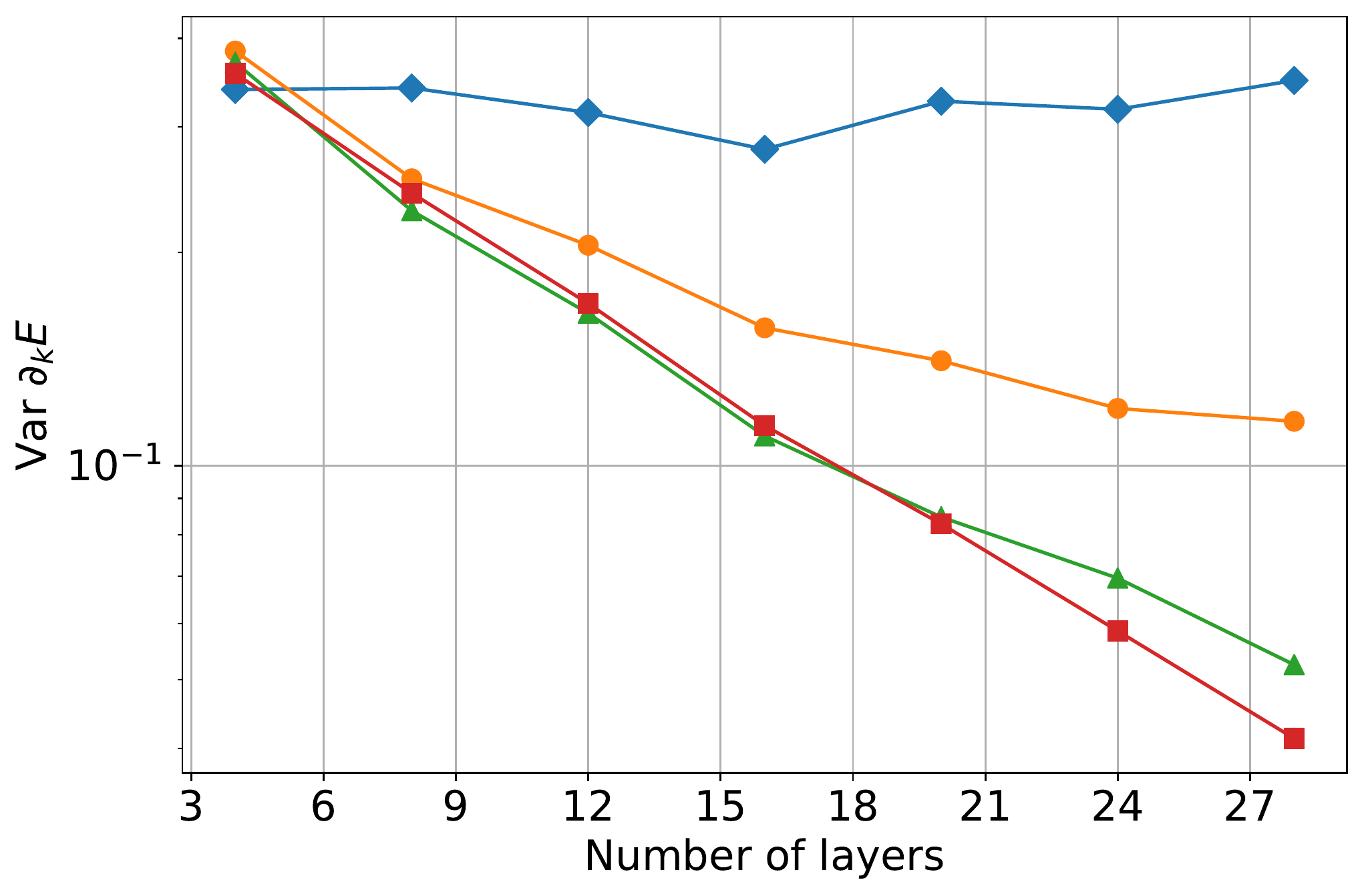}
\caption{Barren plateau effect for the Hamiltonian of Eq.~(\ref{eq:hubbard_nnn}) with $V_1=2t$ and $V_2=t$ depending on the number of qubits as performed by means of Bravyi--Kitaev transformation.}
\label{fig:bk_hubbard}
\end{figure}{}

In the case of one-dimensional interacting spinless fermions with the Hamiltonian (\ref{eq:hubbard_nnn}), the variance both inside and outside the Luttinger liquid regime, mapped onto qubit space by Jordan-Wigner and Bravyi-Kitaev transformations, are, respectively, shown in Figs.~\ref{fig:jw_hubbard} and \ref{fig:bk_hubbard}. In the latter case, the two-qubit gates (\ref{eq:two-qubit_gate}) no longer preserve particle numbers, thus enforcing the use of a more generic ansatz as described below. Surprisingly enough, we do not observe substantial difference in behavior of gradient variance when the model parameters are changed. The fermion-to-qubit mapping, however, impacted the results significantly. For Jordan--Wigner mapping, the exponential dependence on the qubit number emerges right away regardless of the layer number. For Bravyi--Kitaev mapping, the barren plateau regime sets on more gradually. 

\begin{figure}
    \centering
    \includegraphics[width=0.8\linewidth]{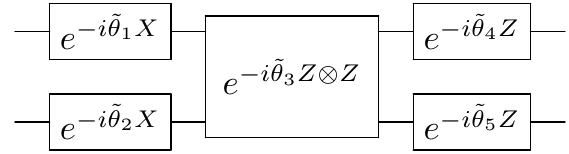}
    \caption{A two-qubit entangler gate consisting of a two-qubit rotation parametrized by a two $\tilde{\theta_3}$, sandwiched between one-qubit rotations, parametrized by $\tilde{\theta_1}$, $\tilde{\theta_2}$, $\tilde{\theta_4}$, and $\tilde{\theta_5}$.}
    \label{fig:entangler}
\end{figure}

\begin{figure}
\centering
\includegraphics[width=\linewidth]{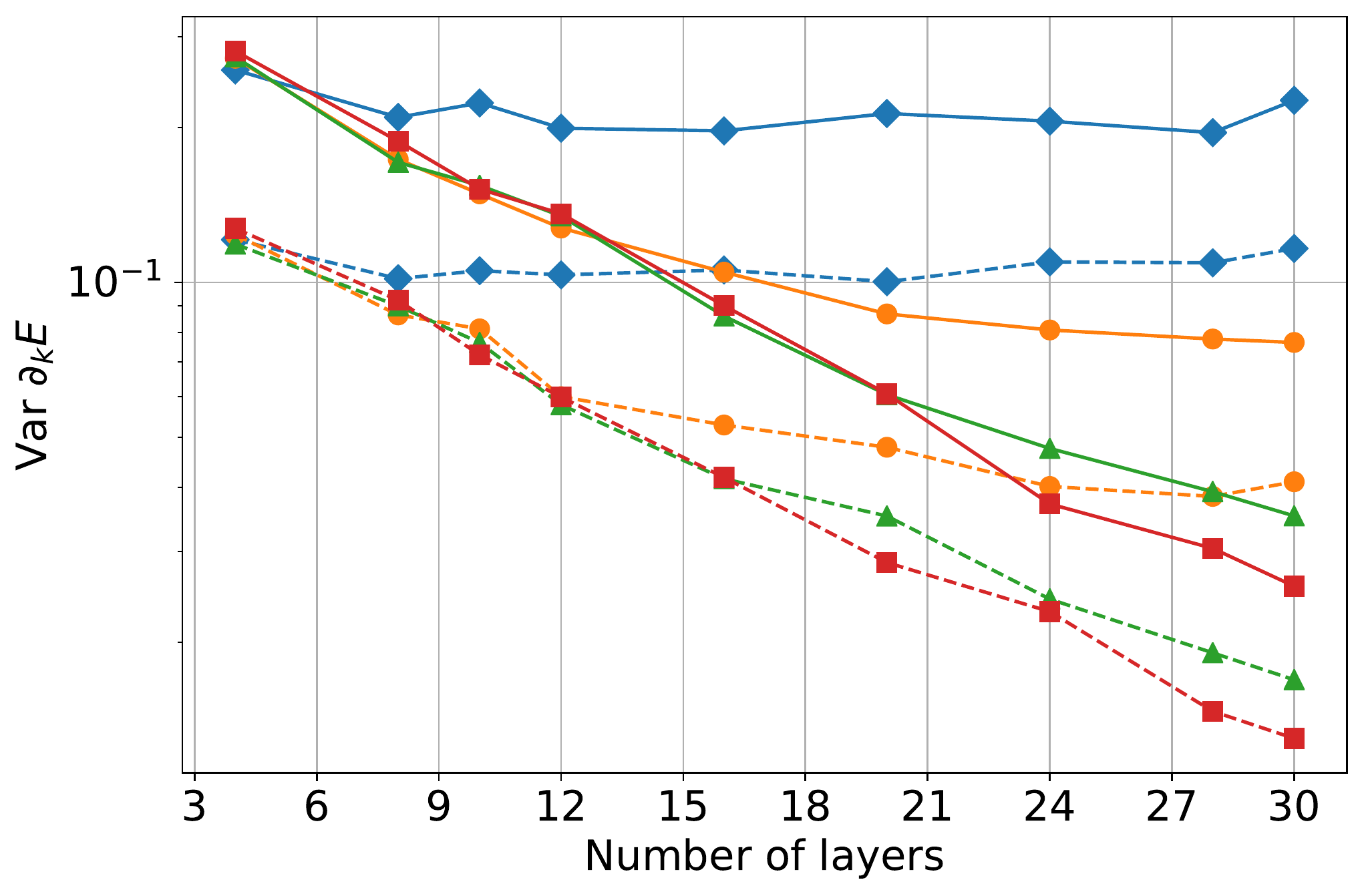}
\caption{Barren plateau effect for the transverse field Ising model of Eq.~(\ref{eq:tfim}) away from criticality with $h=0.1$ (dashed lines) and at the critical point $h=1$ (solid lines).}
\label{fig:tfim}
\end{figure}

To highlight the difference with the case when no fermion mappings are used we consider how the barren plateaus arise in the Ising model with transverse magnetic field $h$:
\begin{equation}
    \label{eq:tfim}
    H_\mathrm{TFIM} = \sum\limits_i Z_i Z_{i+1} + h \sum\limits_i X_i.
\end{equation}
For this model we use a simpler two-qubit entangler gate shown in Fig.~\ref{fig:entangler}, owing to the fact that no fermion mappings are used and particle conservation is irrelevant. Note that the same two-qubit gate has been used for the study of plateaus in case of one-dimensional spinless fermions with next-nearest-neighbor interactions encoded by Bravyi--Kitaev transformation. The gradient behavior away from the critical point ($h = 0.1$) and at the critical point ($h = 1$) is shown in Fig.~\ref{fig:tfim}. In this case, the gradient variance decays exponentially with the number of layers until reaching the plateau regime for the particular number of qubits. Thus, for four qubits the plateau is reached right away, while for 10 qubits 30 layers of the ansatz are still a number belonging to the transition regime. In the meantime, the criticality of the model does not seem to affect this behavior.

{}

\section{Conclusion} 

The scaling up of resources required for VQE is a crucial matter which decides its usefulness for practical implementation in experimental systems with several dozen qubits. Partially for this reason, a variational ansatz should be devised in way to approximate the exact solution  with the necessary precision. Moreover, a chosen ansatz should possess reasonably good convergence properties to guarantee optimization of the cost function in a realistic setting. In this work, we studied the performance of the VQE when applied to one-dimensional spinless fermions with competing nearest- and next-nearest-neighbor interactions. Apart from considering the convergence with respect to energy, we have also addressed density-density correlation function. We performed numerical tests related to the onset of the barren plateau effect \cite{mcclean_barren_2018}. We discovered that fermion-to-qubit mapping is a key factor in the appearance of this effect. This result is consistent with the findings of Ref.~\cite{cerezo_cost-function-dependent_2020}, which connects the susceptibility to plateau with the locality of the Hamiltonian. Indeed, due to trailing $Z$'s, Jordan--Wigner transformation produces terms which are acting on up to $n$ qubits, while Bravyi--Kitaev transformation returns operators which act on $\mathcal{O}(n\log n)$ qubits each. This implies that using the Bravyi--Kitaev transformation may offer better convergence for problems with larger qubit numbers.

In general, given a VQE solution, one can query the expected values of any observables that can be constructed with a polynomial number of Pauli strings. That is, in VQE one measures the expected values of the operators comprising the Hamiltonian. However, if every qubit can be measured in the $X$, $Y$, or $Z$ basis, we can measure any operator that is a tensor product of local Pauli matrices. By repeatedly preparing the state and measuring in different bases, we can estimate the expected value of the sum of such operators. Since the Pauli strings form a complete basis in the space of Hermitian operators, any operator can be evaluated this way. However, a generic one would have an exponential number of terms, hence the polynomial restriction. For example, it was proposed to make use of $\Delta = \langle H^2 \rangle - \langle H \rangle^2$ in order to see the proximity to some eigenstate \cite{kokail_self-verifying_2019}. Likewise, measurement of operators slightly different from the Hamiltonian was proposed as a part of a VQE variant designed to study the excited states of a molecule \cite{colless_computation_2018}. However, any two- and many-body correlation functions can be constructed this way, opening up an avenue to measure multipartite entanglement which is, however, characterized by nonpolynomial computational complexity \cite{Chiara2018}. If nonetheless quantum entanglement is known, one might hope to be able to reconstruct the collective degrees of freedom by a careful study of entanglement of a wave function in the basis of individual spins/electrons.

\acknowledgments 
We gratefully acknowledge fruitful discussions with M.~I. Katsnelson. The work of A.U. and D.Y. was supported by Russian Foundation for Basic Research Project No.~19-31-90159. J.B. acknowledge support from the research project, {\it Leading Research Center on Quantum Computing} (Agreement No.~014/20). 
Numerical simulations were performed using Qiskit. The Jordan--Wigner and Bravyi--Kitaev transformations were done with the use of OpenFermion. 

\bibliographystyle{apsrev4-2}
\bibliography{main.bbl}

\end{document}